\documentclass[twocolumn,rotate,aps,prl,showpacs,amsmath,nobalancelastpage]{revtex4}
\usepackage{graphicx}
\usepackage{bm,eucal}
\begin{document}
\title{Rayleigh--Taylor turbulence in two dimensions}
\author{Antonio Celani$^{(1)}$, Andrea Mazzino$^{(2,3)}$ and Lara Vozella$^{(2,3)}$}
\affiliation{$^{(1)}$CNRS, INLN, 1361 Route des Lucioles, 06560 Valbonne, 
France\\
$^{(2)}$Dipartimento di Fisica, Universit\`a di Genova, Via Dodecaneso 33, I-16146 Genova, Italy\\
$^{(3)}$CNISM and Istituto Nazionale di Fisica Nucleare - Sezione di Genova, Via Dodecaneso 33, I-16146 Genova, Italy  }
\date{\today}
\begin{abstract}
The first consistent phenomenological theory for two and three 
dimensional Rayleigh--Taylor (RT) turbulence has recently been presented by
Chertkov [Phys. Rev. Lett. {\bf 91} 115001 (2003)].
By means of direct numerical simulations we confirm the spatio/temporal 
prediction of the theory in two dimensions and explore the breakdown 
of the phenomenological description due to intermittency effects.
We show that small-scale statistics of velocity and temperature follow
Bolgiano-Obukhov scaling. At the level of global observables we show that the 
time-dependent Nusselt and Reynolds numbers 
scale as the square root of the Rayleigh number.
These results point to the conclusion that Rayleigh-Taylor turbulence
in two and three dimensions, thanks to the absence of boundaries,
provides a natural physical realization of
the Kraichnan scaling regime hitherto associated with the elusive
``ultimate state of thermal convection''.
\end{abstract}

\maketitle 

The Rayleigh--Taylor (RT) instability is a well-known  fluid-mixing
mechanism occuring when a light fluid is
accelerated into a heavy fluid.
For a fluid in a gravitational field, such a mechanism
was first discovered by Lord Rayleigh in the 1880s \cite{ra80}  
and later applied
to all accelerated fluids by Sir Geoffrey Taylor in 1950 \cite{ta50}.\\
RT instability plays a crucial role in many field of science
and technology. As an example, large-scale mixing in the ejecta of 
a supernova explosion can be explained as
a combination of the Rayleigh-Taylor and Kelvin-Helmholtz instabilities
\cite{gull75}.
RT instability also plays a crucial role
in inertial confinement fusion as it finally causes
fuel-pusher mixing that potentially
quenches thermonuclear ignition.
Suppression of the RT instability is thus very crucial for
the ultimate goal of inertial fusion energy.
The final stage of RT instability necessarily leads to the so-called
Rayleigh--Taylor turbulence, the main subject of the present Letter.
Despite the long history of RT turbulence, 
a consistent phenomenological theory 
has been presented only very recently by Chertkov \cite{che03}.
Different behaviors are expected for the 3D and the 2D case.
About the former regime, the ``$5/3$''-Kolmogorov scenario \cite{K41} 
is predicted, while the Bolgiano
picture \cite{bo59} is expected for the 2D case.
This Letter presents the first attempt to compare numerical
results with such phenomenological theory. 
We show that: 
{\it (i)\/} low-order statistics of temperature and velocity
follow Bolgiano scaling;
{\it (ii)\/} there are strong corrections (intermittency)
for higher-order temperature statistics; 
{\it (iii)\/} the behavior of time-dependent global quantities such as
the Nusselt and Reynolds number as a function of Rayleigh number follows
Kraichnan scaling.

\begin{figure}
\includegraphics[scale=0.6]{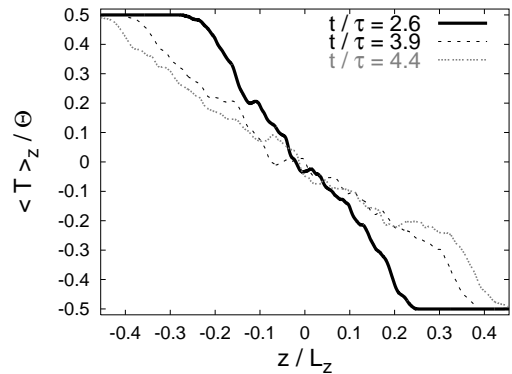} 
\caption{The horizontally  ensemble-averaged temperature field at 
$t/\tau=2.6$ (solid line),  $t/\tau=3.9$ (long-dashed
 line) and at $t/ \tau=4.4$ (short-dashed line). 
The dimensional group $\tau\equiv(L_{z}/(Ag))^{1/2}$ is a
characteristic time scale of the flow.~Here, $A$ is 
the Atwood number, $g$ the gravitational acceleration and 
$L_{z}$ is the vertical size of the box.~The value of $Ag$ is $0.15$.~
Note the almost linear behaviour of $\langle T \rangle_{z}$ 
in the mixed layer.}
\label{fig:1}
\end{figure}

\begin{figure}
\includegraphics[scale=0.4]{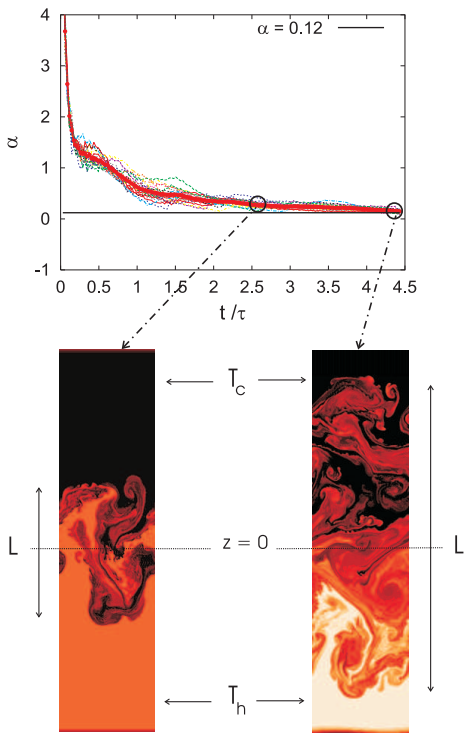}
\caption{The time evolution of the  layer-growth parameter 
$\alpha \equiv [(1/Ag)]d L/d (t^2)$,~$L$ being the mixing layer 
width. Heavy line represents  the ensemble-averaged quantity while
thin lines refer to each individual realization thus giving an indication
of the level of fluctuations.
The horizontal line refers to the value of $\alpha$
found in~\cite{clark} and it is shown for comparison.
The mixing layer width  $L(t)$ 
is defined as 
the distance between $z$-levels  at which
$\mathcal{F}\equiv(\langle T \rangle_{z}-T_{c})/(T_{h}-T_{c})=1\%$ and 
$99\%$, respectively. 
$T_{c}$ and $T_{h}$ are the temperature of the cold and hot fluid, 
before the mixing process takes place.  
Two snapshots of the temperature field are shown at $t/\tau=2.6$ 
(on the left) and at $t/\tau=4.4$ (on the right). 
Dark (white) areas identify cold (hot) regions.}
\label{fig:2}
\end{figure}

The equations ruling the fluid evolution
in the 2D Boussinesq approximation are:
\begin{eqnarray}
\partial_t T + \bm{v}\cdot\nabla T &=& \kappa\Delta T \;\;, 
\label{rt1}\\
\partial_t \omega + \bm{v}\cdot\nabla \omega &=& \nu\Delta \omega -\beta\nabla T\times \bm{g}\;\; ,
\label{rt2}
\end{eqnarray}
T being the temperature field, $\omega={\bm \nabla} \times {\bm v}$ the vorticity, 
$\bm{g}$ the gravitational acceleration, $\beta $ the thermal expansion 
coefficient, $\kappa$ molecular diffusivity and $\nu$ viscosity.\\
At time $t=0$, the system is at rest with
the colder fluid placed above the hotter one.
This amounts to assuming a step function for the initial tempertaure
profile: $T(0,\bm{x})=-sgn(z)\Theta / 2$,
$\Theta$ being the  initial temperature jump.
At sufficiently long times a mixing layer of width $L(t)$ sets 
in, giving rise to a fully developed, nonstationary, turbulent zone,
growing in time as $L(t)\sim t^2$ (i.e., with velocity $u_L(t)\sim t$). 
If, on one hand, 
there is a general consensus
on this quadratic law, which also has a simple physical meaning in terms
of gravitational fall and rise of thermal plumes, 
on the other hand
the value of the prefactor and its possible universality is still
a much debated issue (see, e.g.~Ref.~\cite{clark} and references therein).\\
The statistics of velocity and temperature fluctuations
inside the mixing zone is the realm of application of the pheomenological 
theory of Ref.~\cite{che03}. 
Let us briefly recall the main predictions of this theory and some of its
merits and intrinsic limitations.
The cornerstone of the theory is the quasi-equilibrium picture
where small scales adjust
adiabatically as temperature and velocity fluctuations decay in time. 
Upon assuming that
the temperature behaves as a passive scalar,
the analysis of two-dimensional Navier-Stokes turbulence
leads to two scenarios. While temperature variance 
flows to small scales at a constant flux, the velocity field either
undergoes an inverse cascade with an 
inertial range characterized
by a backward scale-independent energy flux 
or it develops a direct enstrophy cascade
(for background information on two-dimensional turbulence see Ref.~\cite{KM80}
for a theoretical introduction and Refs.~\cite{PT,KG} for experimentally oriented reviews).
Both possibilities actually turn out to be inconsistent \cite{che03}. \\
This apparent deadlock can be broken by rejecting the initial assumption
that temperature behaves as a passively transported quantity at all scales.
Indeed, Chertkov suggests that buoyancy and nonlinear terms
in Eq.~(\ref{rt2}) must be in equilibrium. This is the essence of the 
Bolgiano regime \cite{bo59}, and under
the assumption that temperature fluctuations cascade 
to small scales at a constant rate
one arrives \cite{che03} to the Bolgiano scaling relations:
\begin{eqnarray}
\delta_r T & \sim \Theta\left(\frac{r}{L(t)}\right)^{1/5}\sim (\beta g )^{-1/5}\Theta^{4/5} 
r^{1/5} t^{-2/5}
\label{det} \\
\delta_r v & \sim u_L(t) \left(\frac{r}{L(t)}\right)^{3/5}\sim (\beta g\Theta )^{2/5} 
r^{3/5} t^{-1/5}\;\;\; .
\label{dev}
\end{eqnarray}
The above results constitute a set of mean field (i.e.~dimensional)
predictions which need to  be verified against numerical simulations 
and/or experiments.
Our aim here is to shed some light on both the theory proposed
by Chertkov and to expose the presence of intermittent phenomena
(that could not be addressed within the phenomenological framework of 
Ref.~\cite{che03}) by means of
direct numerical simulations of 
equations (\ref{rt1}), (\ref{rt2}).\\
The integration of both equations is performed by a 
standard $2/3$-dealiased pseudospectral method on a doubly periodic 
domain of horizontal/vertical aspect ratio $L_x/L_z=1/4$. 
The resolution is 
$128\times 4096$ collocation points. 
Different aspect ratios (up to $1:8$) and resolutions (up to $128\times 8192$)
did not show substantial  modifications on the results.
In order to avoid possible inertial range 
contaminations,  no  hyperviscosity/hyperdiffusivity have been used.
The time evolution is implemented by a standard
second-order Runge--Kutta scheme. The integration starts from an
initial condition corresponding to a zero velocity field and to 
a step function for the temperature. Given that the system is intrinsically
nonstationary, averages to compute statistical observables 
are performed over different realizations (about $40$ in the present
study).  The latter are produced by generating initial interfaces 
with sinusoidal waves
of equal amplitude and random phases \cite{clark}. Each realization
is advanced in time until the mixing layer invades the $75\%$ of the
domain.  
The horizontally ensemble-averaged temperature field at three different 
instants is shown in Fig.~\ref{fig:1}. 
It is worth noticing 
the almost linear behavior of the averaged temperature within the mixed 
layer. This is a first clue suggesting a possible relation between
RT turbulence and the 2D Boussinesq driven convection
studied in Refs.~\cite{cmv01,cmmv02}. Further evidences 
will be given momentarily.
In that particular instance of two-dimensional convection,
turbulent fluctuations are driven  
by an external, linearly behaving with the elevation, temperature profile
and the emergence of the Bolgiano  regime clearly appears from data 
\cite{cmv01}. We will argue that 2D RT turbulence corresponds to
the case driven by a linear temperature profile with a mean gradient
that adiabatically 
decreases in time as $\Theta/L \sim t^{-2}$.\\
The mixing layer growth rate is shown in Fig.~\ref{fig:2} in terms
of the growth-rate parameter $\alpha$. Consistently with previous findings
the mixing layer grows quadratically in time,
i.e.~$\alpha$ becomes almost constant in time \cite{nota1}, reaching the value
$\alpha\sim 0.12$.
The latter value is in agrement with the one  found in 
Ref.~\cite{clark} (see the case $\chi=0.01$).

\begin{figure}
 \includegraphics[scale=0.53]{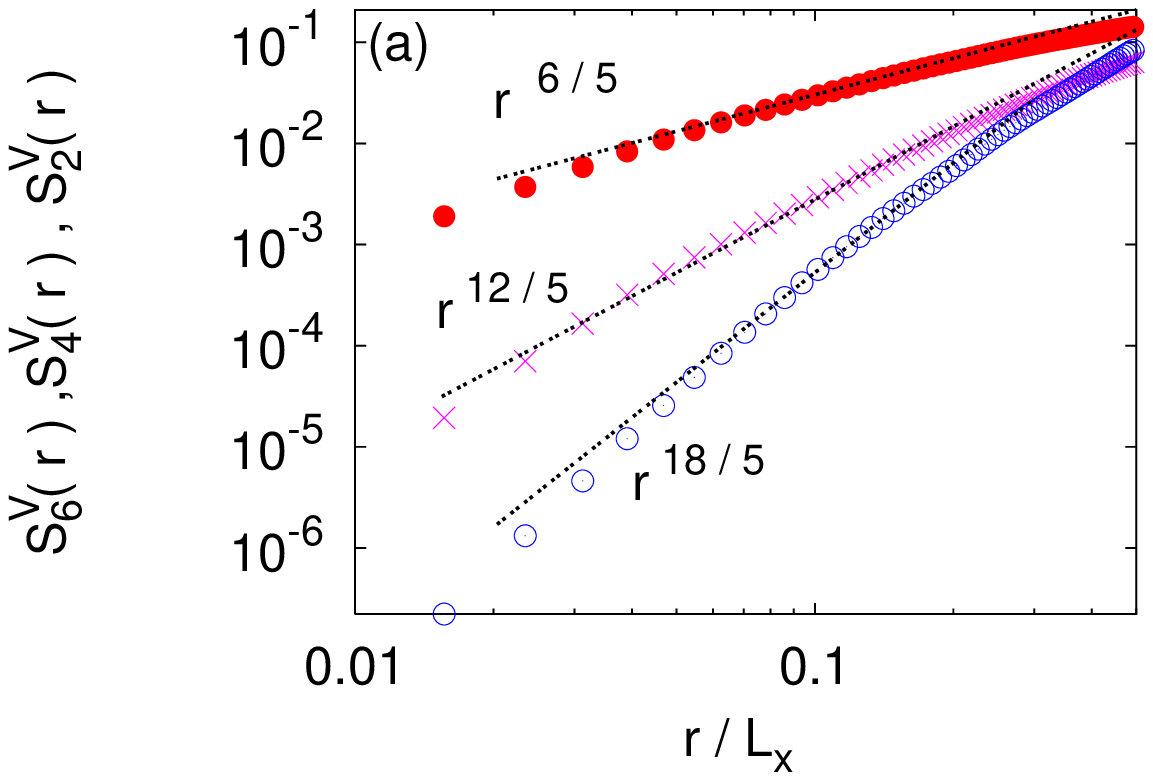}\\ 
 \includegraphics[scale=0.53]{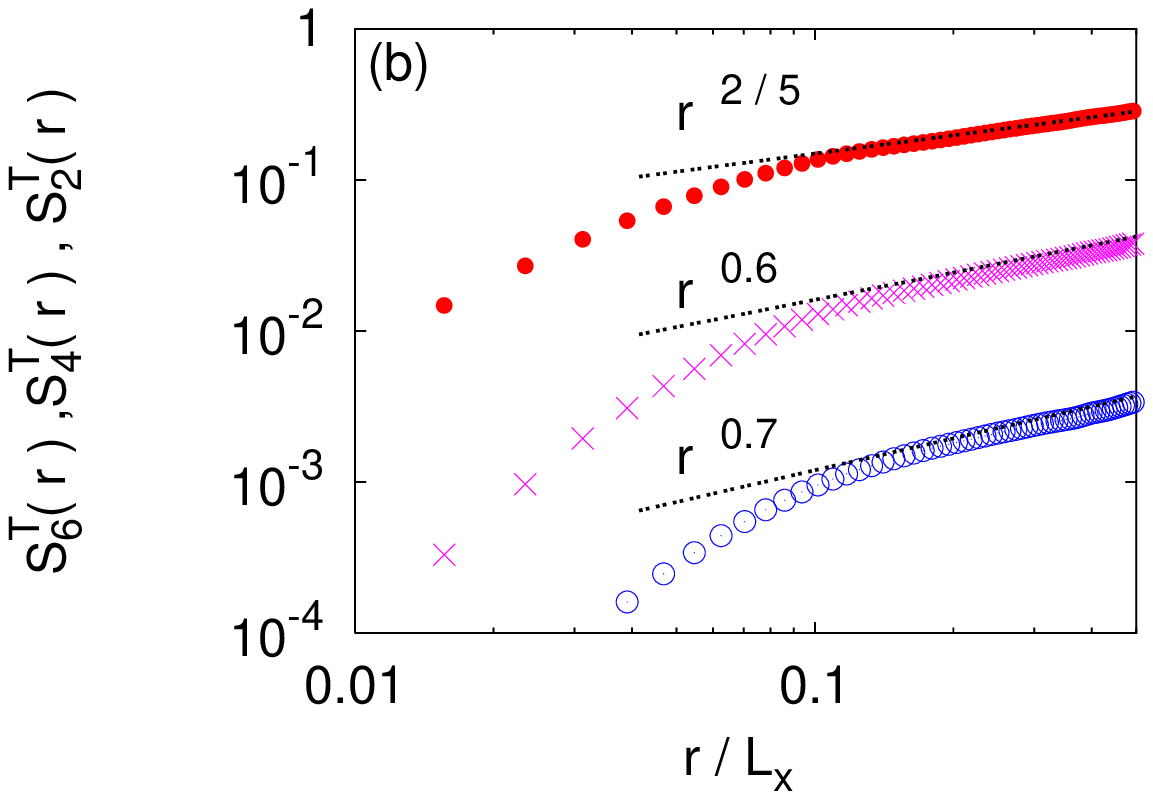} 
\caption{The moments of longitudinal velocity differences $S^{V}_{n}(r)$
(a) 
and temperature  differences $S^{T}_{n}(r)$ (b).
We consider the isotropic contribution to the statistics, by averaging 
over all directions of separations $\mathbf{r}$. In~(a) dashed lines are 
the Bolgiano's dimensional predictions, $S^{V}_{n}\sim r^{3 n/5}$~\cite{bo59}. 
In~(b), for $n=2$ the dashed line is the Bolgiano's 
dimensional prediction:  $S^{T}_{2}\sim r^{2/5}$~\cite{bo59}. For $n=4,6$ 
scaling exponents are 
anomalous, and their values are compatible with 
those of the $2D$ turbulent convection model of Refs.~\cite{cmv01,cmmv02} 
(dashed lines). Moments are averaged over different $L$'s ranging
from $L/L_z=0.4$ to $L/L_z=0.6$ (Fig.~\ref{fig:6}).
}
\label{fig:4}
\end{figure}

In order to quantitatively assess the presence of 
Bolgiano's regime for RT turbulence, let us  focus on scaling behavior
of stucture functions.  For both velocity and temperature differences
they are shown in Fig.~\ref{fig:4}.  On the one hand, 
 second-order structure functions follow the dimensional predictions
(\ref{det}) and (\ref{dev}). On the other hand, 
moments of order $4$ and $6$ display
intermittency corrections for the temperature, e.~g. deviations from
(\ref{det}),  while this is not the case 
for the velocity which shows a close-to-Gaussian probability 
density function  for
inertial range increments (not shown).
The slopes of Fig.~\ref{fig:4} we have associated to $S_4^T$ and $S_6^T$
are relative to the scaling exponents found in Ref.~\cite{cmv01}. This
is a further quantitative evidence in favour of the
equivalence between RT turbulence and Boussinesq turbulence in two
dimensions~\cite{cmv01}. With the present statistics, moments of order higher 
than $6$ are
not accessible. 
\begin{figure}
\includegraphics[scale=0.53]{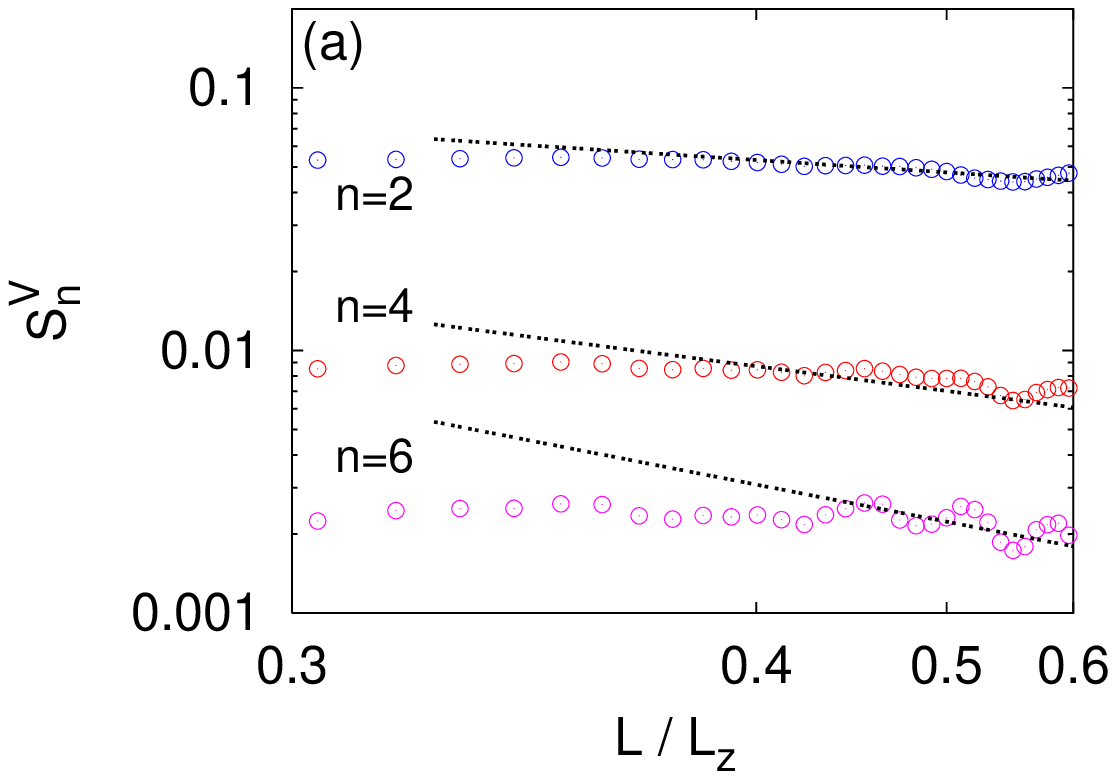}\\ 
\includegraphics[scale=0.53]{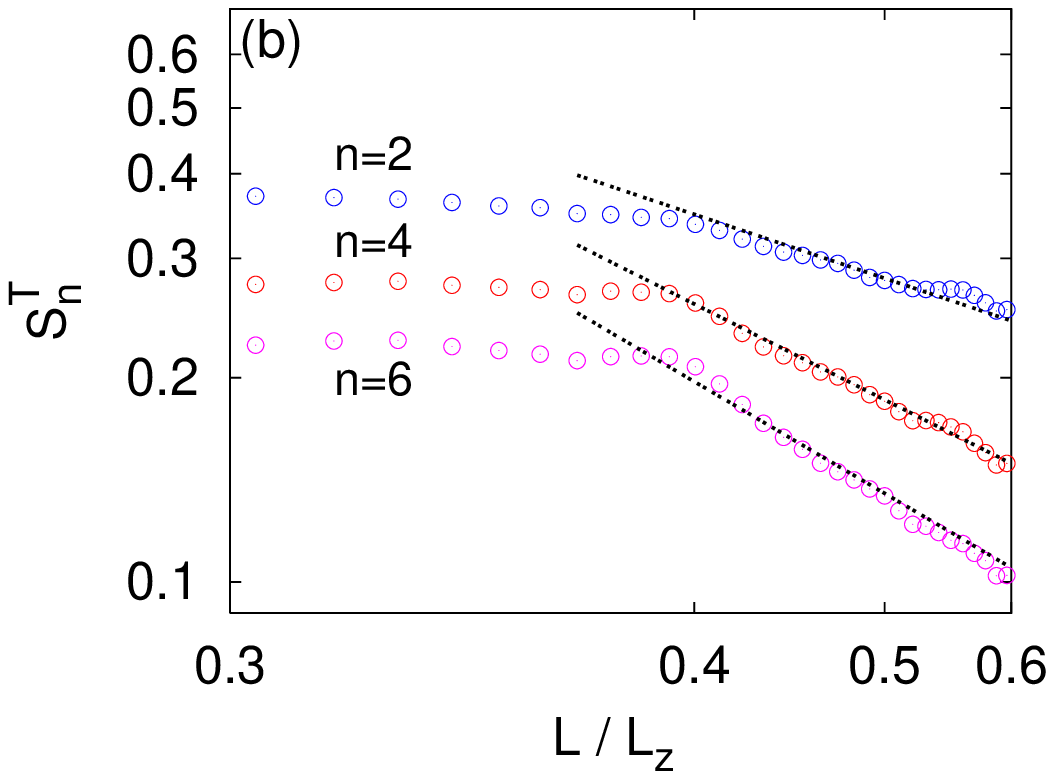} 
\caption{Evolution with 
$L$ of (a) moments of  velocity differences $S^{V}_{n}$
and (b) temperature  differences $S^{T}_{n}$. 
The results are obtained by  averaging 
over a fixed range of separations belonging to the inertial scales. 
We found a better scaling behavior by displaying the behavior of 
structure functions 
as a function 
of the mixing layer width, $L$, rather than with respect to time.  
Those are connected
by the relation $L\sim t^2$.
The dash-dotted lines correspond to the dimensional scaling for velocity
and to the intermittency-corrected scaling for temperature (see text 
for details).}
\label{fig:6}
\end{figure}
We can further corroborate our claim on the possible equivalence
 between RT and driven 2D Boussinesq turbulence
by looking at the temporal behavior of structure functions.
Dimensional predictions immediately follow from Eqs.~(\ref{det})
and (\ref{dev}). The latter are well verified for all displayed orders
for the velocity field (see Fig.~\ref{fig:6}). 
For the temperature field, anomalous corrections start to appear 
at the fourth order and are of the form  
$S_{n}^T(r)\sim \Theta^{n}(r/L(t))^{n/5}(r/L(t))^{-\sigma_{n}}$.
If we assume (see Fig.\ \ref{fig:4}) 
that the present RT model possesses the same spatial
scaling exponents as those of the model presented in Ref.~\cite{cmv01},
i.e.~$S_4^T(r)\sim r^{0.6}$ and $S_6^T(r)\sim r^{0.7}$
(and thus $\sigma_4=0.2$ and $\sigma_6=0.5$), we immediately 
get a prediction for the exponents relative to the temporal behavior.
We just have to remember that $L(t)\sim t^2$ to obtain
the scaling relations $S_4\sim L^{-0.6}\sim t^{-1.2}$ and   
$S_6\sim L^{-0.7}\sim t^{-1.4}$. The latter are compatible with our
results presented in Fig.~\ref{fig:6} (b).\\

\begin{figure}
\includegraphics[scale=0.53]{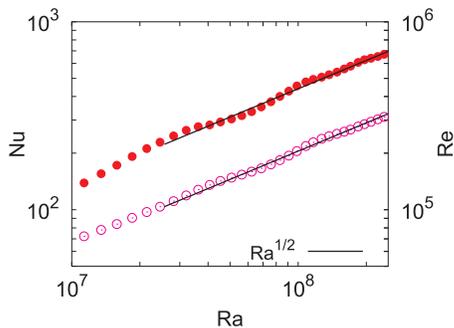}
\caption{The behavior of the Nusselt number (bullets) and Reynolds number
(circles)
versus the
Rayleigh number. 
}
\label{fig:nus}
\end{figure}

We end up by discussing the behavior of turbulent heat flux,
mean temperature gradient and root-mean-square velocity as a function of time.
These quantities are customarily represented in adimensional
variables by the Nusselt number 
$Nu=[\langle v_z T \rangle L / (\kappa \Theta)] + 1$,
the Rayleigh number $Ra= g \beta \Theta L^3 / (\nu \kappa)$
and the Reynolds number $Re=u_{rms} L / \nu$.
 The question about what functional relations exist among these quantities 
is a long-debated issue in the context of three-dimensional Rayleigh-Be\'nard 
turbulence
(see, for example, Refs.\cite{S94,GL00,NSSD00,XBA00,NA03} and citations 
therein). In 2D RT turbulence
exact expressions linking these adimensional numbers to temperature and kinetic
energy input/dissipation rates 
can be derived closely following Ref.~\cite{GL00}. For the RT
case we have to deal 
with the additional complication of time-dependence
yet compensated by the simplification originating from the absence 
of boundary effects.
These relations read $\partial_t \langle {\bm v}^2/2 \rangle = 
\nu \kappa^2 L^{-4} (Nu-1) Ra - \epsilon_v $ and $\partial_t \langle \theta^2/2 \rangle
= \kappa \Theta^2 L^{-2} (Nu-1) - \epsilon_{\theta}$, $\theta$ being the 
departure from the mean temperature profile. 
Since in 2D RT kinetic energy is 
transferred upscale we have a negligible $\epsilon_v$ and we can estimate 
a rate-of-change of kinetic energy  $(\beta g \Theta)^2 t$ from 
 Eq.~(\ref{dev}) and $L= \beta g \Theta t^2$. For the temperature we have
$\langle \theta^2 \rangle \sim \Theta^2$ independent of time. Temperature 
performs a direct cascade and thus
dissipation can be estimated as $\delta_L \theta^2 \delta_L v / L \sim 
\Theta^2 / t$. Plugging those estimates into the exact relations yields 
$Nu \sim (\beta g \Theta)^2 \kappa^{-1} t^3$ and 
$Ra \sim (\beta g \Theta)^4 (\nu \kappa)^{-1} t^6$, therefore 
$Nu \sim Ra^{1/2} Pr^{1/2}$.
As for the Reynolds number, Eq.~(\ref{dev}) gives 
$Re \sim (\beta g \Theta)^2 \nu^{-1} t^3 \sim Ra^{1/2} Pr^{-1/2}$. 
It is worth remarking that a similar analysis for the 3D RT
case leads {\em mutatis mutandis} to the same scaling laws.
These, when considered as a function of $Ra$, coincide with the results derived
by Kraichnan  for the pure bulk contribution to 
3D Rayleigh-B\'enard turbulence.
However, such Kraichnan scaling regime for RB convection, also dubbed ``the
ultimate state of thermal convection'', has so far eluded both experimental and
numerical confirmation \cite{Amati2005}.  Additionally, it has been shown that
it is not realizable in the analytically tractable case of $Pr$ going to
infinity \cite{Doering2005}. The reason may be traced back to the fundamental
role played by the boundaries in establishing the turbulent heat transport in
Rayleigh-B\'enard convection.  Indeed, when boundaries are artificially 
removed as in
the numerical simulations of Refs.~\cite{LT03,CLTT05}, the Kraichnan scaling is
clearly observed. In this context, Rayleigh-Taylor turbulence provides a
natural framework where heat transport takes place exclusively by bulk
mechanisms and thus provides a physically realizable example of the Kraichnan
scaling regime, inviting further experimental and numerical effort in this
direction.  

\vspace{2mm}
\noindent AM and LV have been supported by COFIN 2005 project 
n.~2005027808 and by CINFAI consortium.
Simulations have been performed at CINECA (INFM parallel
computing initiative). AC acknowledges the support of the 
European Union under the contract HPRN-CT-2002-00300 and PICS-CNRS 3057.
Useful discussions with M.~Chertkov are gratefully acknowledged.


\begin{thebibliography}{99}
\bibitem{nota1}
We actually observe a slight decrease in time which has also been
detected in Ref.~\cite{clark} and whose origin still
remains a subject of discussions.
\end{thebibliography}
\end{document}